\documentstyle[12pt]{article}
\renewcommand{\d}{\delta}

\newcommand{\mod}{{\rm mod}}

\newcommand{\WH}{{\rm WH}}
\newcommand{\QFT}{{\rm QFT}}
\newcommand{\U}{{\rm U}}
\newcommand{\D}{\Delta}

\newcommand{\e}{\epsilon}
\newcommand{\ep}{\varepsilon}

\newcommand{\ar}{\longrightarrow}
\newcommand{\w}{\omega}

\newcommand{\la}{\lambda}
\renewcommand{\a}{\alpha}
\begin{document}
\title{How behavior of systems with sparse spectrum can be predicted on a 
quantum computer}
\author{Yuri Ozhigov
\thanks{This work was performed while the author was at Bell Labs, Lucent
Technologies. It was supported by the U.S. Army Research Office 
under contract no. DAAG55-98-C-0040;\ 
 e-mail addresses:   y@oz.msk.ru, ozhigov@hotmail.com}}
\date{}
\maketitle
\begin{abstract} Call a spectrum of Hamiltonian sparse if each eigenvalue 
can be quickly restored with accuracy $\varepsilon$ from its rough 
approximation in within $\varepsilon_1$ by means of some classical 
algorithm. It is shown how a behavior of system with sparse spectrum
up to time $T=\frac{1-\rho}{14\varepsilon}$ can be predicted with fidelity 
$\rho$ on quantum computer in time $t=\frac{4}{(1-\rho)\varepsilon_1}$ plus the 
time of classical algorithm. The quantum knowledge of Hamiltonian $H$ 
eigenvalues is considered as a wizard Hamiltonian $W_H$ which action on
 any eigenvector of $H$ gives the corresponding eigenvalue. Speedup of 
evolution for systems with sparse spectrum is possible because for such
 systems wizard Hamiltonians can be quickly simulated on a quantum computer. 
This simulation, generalizing Shor trick, is a part of presented 
algorithm. 
In general case the action of wizard Hamiltonian cannot be simulated in time 
smaller than the dimension of main space which is exponential of the size of 
quantum system. For an arbitrary system (even for classical) its behavior 
cannot be predicted on quantum computer even for one step ahead. This method 
can be used also for restoration of a state of an arbitrary primary system in
 time instant $-T$ in the past with the same fidelity which requires the same 
time. 
\end{abstract}

\section{Introduction and background}

A behavior of typical quantum system cannot be analyzed at hand or by 
classical computer because huge dimension of its Hamiltonian makes it 
impossible to solve Shr\"oedinger equation even numerically. Nevertheless
 one can expect that this analysis would be easier in the framework of 
quantum computing. The idea of this approach is to force one quantum system 
to simulate a behavior of other more simple primary system with some profit 
in time. In the particular case where the primary system is a classical 
computer with oracle this is a problem of quantum speedup for classical
computations: given a classical algorithm with oracle, is there a quantum
 algorithm computing the same function faster using the same oracle quantumly?
 Examples: Shor factoring algorithm (look at \cite{Sh}), Grover search 
algorithm (look at \cite{Gr}).

In general case the problem of prediction has the following form: given a 
low of evolution for some system (classical or quantum) is there a device 
(wizard) of not exponential size which can predict the behavior of this 
system? To predict the behavior means that given a time instant, wizard 
returns a state of initial system in this instant earlier than this state 
appears in its natural evolution. Note that in general case of quantum 
evolution a wizard may use quantumly not only Hamiltonian (like in case
 of quantum computations with oracle) but some hidden information about
 primary system. What kind of such information may be useful for prediction? 
This is the information about eigenvalues of primary Hamiltonian. This article
 analyses the possibility of quantum predictions in terms of spectrum of
 primary system. 

Quantum method of finding eigenvalues was first presented by Abrams and
 Lloyd in their work \cite{AL}. Their method, generalizing a quantum part of 
Shor factoring algorithm, uses Quantum Fourier transform 
(QFT) and requires of order $N$ implementations of initial Hamiltonian where
 $N$ is the dimension of the main space. Thus, their algorithm requires 
exponential time and it cannot be used for quantum speedup. However, for 
the systems whose spectrum is sparse the idea of this approach with QFT may 
be used for predictions. I proceed with the exact definitions. 

\section{Definitions}

Let $H$ be some Hamiltonian which induces unitary transformations $U$ in some 
$N$ dimensional space $Y$ of states of $n$ qubits by the usual rule in quantum
 mechanics: $U(\tau)=\exp(-iH\tau /h)$, $N=2^n$. In order to apply notions 
from the algorithm theory, such as complexity, etc., we assume that the time 
of Hamiltonian action is always $\tau =1$. We may of course choose another 
value for $\tau$ but this would only involve the change in time scale without 
any subsequences for the algorithm complexity. 
Call a system primary if its evolution is determined by unitary transformation
 $U$.

We assume that the complexity of computation is the number of $U$
 applications. Time of all other transformations we use here is assumed
 to be negligible comparatively with the time of computation. Quantum Fourier
 transform (QFT) requires the time of order $n^2$ (look at \cite{Sh}). Here we 
shall use this transform for $\log_2 (M)$ gubits in the following form:
$$
\QFT_M :\ \ |s\rangle \ar \frac{1}{\sqrt{M}}\sum\limits_{l=0}^{M-1} \exp 
(-2\pi i s l/M)|l\rangle
$$

Let eigenvalues of $H$ have the form $-2\pi\w_k h\ ,k=0,1,\ldots ,N-1$, when 
frequencies $\w_k$ are real numbers from the segment $[0,1)$. Then the 
eigenvalues of $U$ will be $\exp(2\pi i\w_k )$. This is not loss in generality 
because 
we always can choose other unit $\tau$ for Hamiltonian action time.  Let us 
give the precise definitions of the "knowledge of eigenvalues" of $H$. It
 means that there exists another Hamiltonian $W_H$ called a wizard for $H$
which acts in $N^2$ dimensional Hilbert space and returns an approximation 
of frequencies $\w_k$ in within $1/N$ given eigenvector $\Phi_k$ of $H$. 
Introduce the following notations. 
Every frequency $\w_k$ has binary notation $0.\e_1 \e_2 \ldots$.
Let $p\leq n$ be some integer, $M=2^p$.
Denote the string of ones and zeroes $\e_1 \e_2 \ldots e_p$ by $\bar\e_k^p$.
 $\e_k^p$ may be considered as integer if we cancel all first zeroes. And vise 
versa, every integer $l$ less than $M$ can be written in form $\bar\e_k^p$ if
 we add suitable number of zeroes in front. Then the number $\tilde\w_k^M =
 0.\bar\e_k^p$ will be approximation of $w_k$ in within $1/M$. We shall 
define
this number $0.\bar\e_k^p$ from $[0,1)$ by $(0.l)_p$. Wizard action in these 
notations will be
$$
W_U |\Phi_k ,b \rangle \ar |\Phi_k , b\bigoplus\bar\e_k^p \rangle
$$
where $\bigoplus$ is the biwise addition modulo $2$. 

At last introduce auxiliary transformations $\WH$ and $\U^p_{seq}$. Let the 
memory is divided into the main part $x$ of $n$ qubits and ancilla $a$ of
 $p\leq n$ qubits: $|x,a\rangle$, $M=2^p$. Put:

1). $\WH |x,a\rangle = \frac{1}{\sqrt{M}} \sum\limits_{s=0}^{M-1} 
(-1)^{a\cdot s} |x,s\rangle$. This is Walsh - Hadamard transform applied to
 ancilla where $\cdot$ denotes dot product modulo 2.

2). $\U^p_{seq} |x,a\rangle = | \U^a x,a\rangle$. This is the result of $a$ 
sequential applications of $U$ to the main register.

Walsh-Hadamard transform can be fulfilled in a standard model of quantum 
computer. To fulfill $\U^p_{seq}$ it would suffice to use the following 
oracle $\U_{cond}$ (conditional application of $\U$) depending on $\U$:
$\U_{cond} |x,\a\rangle\ar|x',\a\rangle$, where $x'=x$ if $\a=0$ and $x'
=\U x$ if $\a =1$, $\a$ is one qubit register. An application of $\U_{cond}$
 cannot be reduced to the simple using of $\U$ as an oracle because a
 conditional application of $\U$ is quantumly controlled by the second 
register. One proposal about its practical implementation can be found 
in the section 4.5.

\section{Wizard transformation}

\subsection{How wizard predicts evolution}

Assume temporarily that we have exact equations $\w_k =\tilde\w_k^N$. 
Given a wizard transformation $W_U$ how can we predict evolution of 
initial system?  Let $|\xi\rangle$ 
denote initial state as a contents of $n$ qubits main register. Let 
\begin{equation} 
\xi =\sum\limits_{k=0}^{N-1} x_k
|\Phi_k
\label{1}\end{equation}

 be the expansion of our state in basis of eigenvectors of $U$. Add $n$ 
qubits ancillary register initialized by zeroes and obtain the state $|\xi , 
0^n\rangle $. 

Now apply the wizard transformation $W_U$ to the main register ($p=n$). 
It gives the state
$$
\xi' = \sum\limits_{k=0}^{N-1} x_k
|\Phi_k , \bar\e_k^p \rangle .
$$
Given the number $t$ we can turn each state in the first register by angle
 determined by the second register and $t$: $2\pi \w_k t$. This gives the 
state
$$
\xi '' = \sum\limits_{k=0}^{N-1} x_k \exp (2\pi i\w_k t) | \Phi_k , 
\bar\e_k^p \rangle .
$$
At last apply wizard again obtaining
$$
\xi''' = \sum\limits_{k=0}^{N-1} x_k \exp(2\pi i \w_k t)|\Phi_k , 0^n 
\rangle ,
$$
and now cancellation ancilla gives exactly the state $U^t |\xi\rangle$ 
which is the state in time instant $t$ of initial system. Thus given a 
wizard we can predict the behavior of initial system provided the action 
of wizard takes smaller time than $t/2$. 

\subsection{Simulation a wizard}

In this subsection I describe the procedure of simulation a wizard action
 on zero ancilla. The particular case $p=n$ of this procedure was proposed
 in other purpose in the paper \cite{AL}.

We start from the state of the form (1) with ancilla attached initialized 
by zeroes. Wizard action on the initial state with zero ancilla is defined
 by the following:
\begin{equation}
\QFT_M \ \U^p_{seq} \ \WH \ |\xi ,0^p \rangle .
\label{2}
\end{equation}

Why this procedure must work? Let us again assume temporarily that $\w_k 
=\tilde\w_k^{M}$ exactly. Then we have $\WH |\xi , 0^p \rangle = 
\chi_0 = \frac{1}{\sqrt{M}}\sum\limits_{k=0}^{N-1} \sum\limits_{s=0}^{M-1} 
x_k |\Phi_k , s\rangle$. Application of
 $\U^p_{sec}$ gives: $ \chi_1 = \U^p_{sec} \chi_0
=\frac{1}{\sqrt{M}}\sum\limits_{k=0}^{N-1} \sum\limits_{s=0}^{M-1}
x_k \exp(2\pi i\w_k s) |\Phi_k , s\rangle$. Now $\QFT_M \chi_1 =
 \sum\limits_{k=0}^{N-1} x_k
|\Phi_k , \bar\e_k^p \rangle$.
This is just the wizard action. But in general case the result of QFT
transform has not so simple form. It means that we cannot merely clean
ancilla by repetition of wizard action in predicting algorithm even if
we have a wizard action on all words. Hence the precision of a wizard
simulation by (2) must be elaborated in more details.

\subsection{Accuracy of a wizard simulation}

Let $\{\tilde\w_{k,i} \}$ be some set of integers. Denote
$L_\ep (\tilde\w_{k,i}) = \{ i:\  |(0.\tilde\w_{k,i} )_p - \w_k |\leq\ep$
or \newline
$|(0.\tilde\w_{k,i} )_p - \w_k -1|\leq\ep\}$. Let $\xi$ be a
state of the form (1).

{\bf Definition}
{\it A transformation $W$ of the form 
$$
W:\ |\xi , 0^p \rangle \ar\sum\limits_{k=0}^{N-1}\sum\limits_{i=0}^{M-1} 
\la_{i,k} |\Phi_k ,
\tilde\w_{k,i} \rangle
$$
is called a transformation of $W_{\d, \ep}$ type if 
$\ \ \sum\limits_{k=0}^{N-1}\sum\limits_{i\in L_\ep (\tilde\w_{k,i})} 
|\la_{i,k}  
|^2 \geq 1-\d\ \ 
$ for any $\xi$.
}
Thus, $\d$ is an error probability from the quantum superposition, and 
$\ep$ is a precision of eigenvalues approximations.

\newtheorem{Lemma}{Lemma}
\begin{Lemma} The transformation $\QFT_M \ \U^p_{seq} \ \WH \ $ belongs 
to the type $W_{\frac{1}{K},\frac{K}{M}}$ for any $K\in\{ 1,2,\ldots ,M\}$.
\end{Lemma}

{\it Proof}

Denote $\QFT_M \ \U^p_{seq} \ \WH \ |\xi ,0^p \rangle $ by $\chi _{p,2}$. 
Then we have
\begin{equation}
\chi_{p,2} = \frac{1}{M} \sum\limits_{k=0}^{N-1} \sum\limits_{l=0}^{M-1}
 x_k H_{l,k} |\Phi_k , l\rangle
\label{3}
\end{equation}
where $H_{l,k} =\sum\limits_{s=0}^{M-1} \exp (2\pi is(\w_k -(0.l)_p ))$. 
Put $\D =\w_k -(0.l)_p$. Then by summing the progression we obtain 
\begin{equation}
H_{l,k} =\frac{1-\exp (2\pi iM\D)}{1-\exp (2\pi i\D)} .
\label{4}
\end{equation}

Let $\D= \frac{K}{M} +\d ,\ K=K(l)$, where $0\leq\d <1/M,\ K$ integer. 
Then each $H_{l,k}$ depends on $K:\ H_{l,k} =H_{l,k} (K)$. Fix some value
for $K$: $K_0$. It means that the accuracy of eigenvalues approximations 
will be $K_0 /M$. Estimate the sum of squared amplitudes for states 
$|\Phi_k ,l\rangle$ for which $K:\ K_0 \leq K\leq M-K_0$. Using that 
module of denominator in (4) is separated
from zero by $\min \{\frac{\pi K}{M} , \frac{\pi (M-K)}{M} \}$, we have:
$$
\begin{array}{cc}
&\sum\limits_{k=0}^{N-1}\sum\limits_{K=K_0}^{M-K_0} |x_k |^2 \frac{1}{M^2} 
|H_{l,k} (K)|^2 \leq\\
&4\sum\limits_{k=0}^{N-1}\sum\limits_{K=K_0}^{M-K_0}
 |x_k |^2 /\pi^2 K^2 \leq 
\frac{4}{\pi^2} \sum\limits_{K=K_0}^{M-K_0} \frac{1}{K^2} \leq\frac{1}{K_0}.
\end{array}
$$
 This is exactly what is needed. Lemma 1 is proved. 

What would happen if we apply other sequence of transformations
 $\QFT_M^{-1} \  (\U_{seq}^p)^{-1} \ \WH$ instead of $\QFT_M \ \U_{seq}^p
 \ \WH$ for revealing frequencies? 

\begin{Lemma}
Let for some state $\chi$
$$
\begin{array}{ccc}
\chi_2 &= \QFT_M \ \U^p_{seq} \ \WH |\chi ,  0^p \rangle &= 
\sum\limits_k \sum\limits_{l=0}^{M-1} y_{k,l} |\Phi_k ,l\rangle\\
\chi '_2 &= \QFT_M^{-1} \ (\U_{seq}^p )^{-1} \ \WH |\chi ,  0^p \rangle &= 
\sum\limits_k \sum\limits_{l=0}^{M-1} y'_{k,l}|\Phi_k ,l\rangle
\end{array}
$$
Put $\d_{k,l} =\tilde\w_k^N -(0.l)_p$, $\D =\w_k -(0.l)_p$, $\chi''_2 = 
\sum\limits_k \sum\limits_{l=0}^{M-1} y'_{k,l} 
\exp (2\pi i (M-1) \d_{k,l})|\Phi_k ,l\rangle$. Let $M/N <\e$.
Then 
$$
\| \chi''_2 -\chi_2 \| <7\e  
$$
\end{Lemma}

{\it Proof}

We have $y_{k,l} =x_k H_{k,l}$ (look at (3),(4)),
where $H_{k,l} =H_{k,l} (\D)$.
Then $y'_{k,l} =x_k  H_{k,l} (-\D )$. Now $H_{k,l} (\D )=H_{k,l} (-\D )
\exp (2\pi i (M-1)\D )$, $|\d_{k,l} -\D |\leq 1/N$, $2\pi (M-1)|\d_{k,l} 
-\D |\leq 7M/N$ and Lemma 2 follows. 

\subsection{Complexity of a wizard}

It is readily seen that the above procedure for $p=n$ requires $N$ 
applications of the initial transformation $U$, hence it cannot be 
used on purpose to predict its evolution. It turns that in general 
case the following theorem takes place.

\newtheorem{Theorem}{Theorem}
\begin{Theorem}
It is impossible to simulate a wizard action with precision up to $O(1/N)$ 
using less than $\Omega (N)$ conditional application of initial transformation
 $\U_{cond}$.
\end{Theorem}

{\it Proof}

A lower bound for the time of quantum computation of the PARITY function is 
$N/2$ by the results \cite{FGGS} and \cite{BBCMW}. These constructions can be 
extended to the case when we can use more general oracle $\U_{cond}$ instead 
of $\U$ when evaluating PARITY.
Assuming that a wizard action can be simulated using less than $\Omega (N)$ 
application of $\U_{cond}$, in view of the previous subsection we would be able
 to compute the function PARITY on a quantum computer in time less than $N/2$
(adding appropriate constant to the number $n$ in order to enhance accuracy)  
which leads to the contradiction. Theorem 1 is proved.

Call a unitary transformation $U$ classical if it maps basic states to basic
 states. One can ask is there any way to predict a behavior of classical 
system on a quantum computer? This statement is independent of a form of 
spectrum. It turns that in general case a behavior of the bulk of classical 
systems in a short time frame is impossible on quantum computer even on one 
step ahead (look at \cite{Oz}). Here short time means approximately 
$O(N^{1/7})$, where $N$ - the number of all states.

But for one class of systems the quantum prediction is possible. This is 
the class of systems with sparse spectrum.

\section{Prediction of the evolution of the systems with sparse spectrum}

\subsection{Case of precise eigenvalues}

What would happen if we use less number of qubits $p<n$ instead of $n$ in the 
ancilla? Simulation of the wizard will be shorter, but it will be simulated
 with the corresponding loss in precision and we obtain no prediction. 

Nevertheless, if the spectrum is sparse then prediction is possible. 
Indeed, suppose that we have classical algorithm $h$ enhancing the accuracy 
of eigenvalues approximation. 

Namely, let $p<n$, $M=2^p$ and let $h:\ \{ 0,1\}^p \ar \{ 0,1\}^n$ be integer 
function mapping rough frequency approximation up to $1/M$ to the more precise
 approximation up to $1/N$. 

Given the initial state $\xi$ and time instant $t$ we shall now describe the 
procedure of quick finding $U^t \xi$. The idea is simple:
Repeat the procedure from above with only $p$ ancillary qubits instead of $n$,
 and enhance accuracy by $h$. 

At first we assume for the simplicity that we have exact equalities 
$\w_k =\tilde\w_k^M$. Here is the detailed description of the algorithm.

First we apply
Walsh-Hadamard transform to the ancilla and obtain  
$$
\chi_0 = \frac{1}{\sqrt{M}}\sum\limits_{k=0}^{N-1} \sum\limits_{s=0}^{M-1}
 x_k |\Phi_k , s\rangle 
$$

Then we apply $\U_{seq}^p$. Since $s\leq p$ this operation requires $p$ 
conditional applications of $U$. This gives the state 
$$
\chi_1 = \frac{1}{\sqrt{M}}\sum\limits_{k=0}^{N-1} \sum\limits_{s=0}^{M-1} 
x_k U^s |\Phi_k , s\rangle  = \frac{1}{\sqrt{M}}\sum\limits_{k=0}^{N-1}
 \sum\limits_{s=0}^{M-1} \exp(2\pi i\w_k s) x_k |\Phi_k , s\rangle .
$$

Now application of $\QFT_M$ to the second register yields:
$$
\chi_2 = \sum\limits_{k=0}^{N-1} x_k |\phi_k , \tilde \w_k \rangle
$$
where $0.\tilde w_k$ is approximation of eigenvalue $\w_k$ in within $1/M$. 

This is the point when we will use that the spectrum is sparse. Add one more 
register with $n$ qubits initialized by zeroes:
 $\sum\limits_{k=0}^{N-1} x_k |\Phi_k , \tilde \w_k ,0^n \rangle$. 
Apply the unitary version of the algorithm $h$ enhancing accuracy to two 
ancillary registers:
$|a,b\rangle \ar |a,b\bigoplus h(a)\rangle$. This gives the state
$\sum\limits_{k=0}^{N-1} x_k |\Phi_k , \tilde \w_k ,h(\tilde\w_k )\rangle$. 
Now we turn every state of the form $|\Phi_k , \tilde \w_k , l \rangle$ to 
the angle $2\pi \cdot 0.l\cdot t$ and obtain the state 
$\sum\limits_{k=0}^{N-1} x_k \exp(2\pi i\cdot 0.l\cdot t) |\Phi_k ,
 \tilde\w_k , l \rangle$. Here $t$ is a given time instant. Now apply
 unitary version of $h$ which cleans up $n$ last ancillary qubits and 
then discard them:
$$
\sum\limits_{k=0}^{N-1} x_k \exp (2\pi \w_k t)|\Phi_k ,\tilde\w_k \rangle ,
$$
and again $\QFT_M$ to the $p$ ancillary qubits:
$$
 \frac{1}{\sqrt{M}}\sum\limits_{k=0}^{N-1} \sum\limits_{s=0}^{M-1}
 x_k \exp (2\pi i\w_k t)
\exp (-2\pi i \tilde\w_k s)|\Phi_k ,s\rangle ,
$$
then again $\U_{seq}^p$ and Walsh-Hadamard transform to the ancilla. 
The result will be:

$\sum\limits_{k=0}^{N-1} x_k \exp(2\pi i\w_k t) |\Phi_k ,0^{p} \rangle$.
 The wanted state is in the main register now. 

Note that all the work with enhancing an accuracy here is not mandatory because
 $\w_k
=\tilde\w_k^M$ exactly.
Why this scheme doesn't work for the case when $\tilde\w_k^M$ are only 
approximations of the true eigenvalues $\w_k$ ? The point is that $\chi_2$ 
will not have so simple 
form and in addition multiplication frequencies by $t$ will cause big error
 for $t=O(N)$. Thus for the general case more refined algorithm is needed.

\subsection{General case}

We have numbers $n$ and $p$. Choose some $q: \ p\leq q<n$ and put $L=2^q$.
Without loss in generality we can extend $h$ to the mapping 
$h:\ \{ 0,1\}^q \ar\{ 0,1\}^n$ so that if $\tilde\w_k^L =(0.x)_q$ then $
\tilde\w_k^N =(0.h(x))_n$.

Our initial state has the form 
$$
\xi =\sum\limits_{k=0}^{N-1} x_k |\Phi_k ,0^q \rangle
$$

1. Apply  $\QFT_L \U^q_{seq} \WH$ to the initial state to get $\chi_{q,2}$. 

2. Add one more register with $n$ qubits initialized by zeroes: \newline
$ \sum\limits_k \sum\limits_{l=0}^{L-1} y_{k,l} |\Phi_k ,l,0^p \rangle $ and
 apply the unitary version of the algorithm $h$ enhancing accuracy to two 
ancillary registers:
$\chi_3 = \sum\limits_k \sum\limits_{l=0}^{L-1} y_{k,l} |\Phi_k ,l, h(l) 
\rangle $.  

3. Turn every state of the form $|\Phi_k , l ,h(l) \rangle$ to the angle 
$2\pi \cdot 0.h(l)\cdot t$ and obtain the state $\chi_4 = \sum\limits_k 
\sum\limits_{l=0}^{L-1} y_{k,l} \exp(2\pi i\cdot 0.h(l)\cdot t)|\Phi_k ,l,
 h(l)
 \rangle $.  Here $t$ is given time instant. 

4. Put $\d^*_{k,l} =(0.h(l))_n -(0.l)_q$. 
Turn every state of the form $|\Phi_k , l ,h(l) \rangle$ to the angle
$-2\pi (L-1)\d^*_{k,l}$. Denote the result by $\chi_5$.

5. Now apply unitary version of $h$ which cleans up $n$ last ancillary qubits 
and then discard them: 

$\chi_6 = \sum\limits_k \sum\limits_{l=0}^{L-1} y_{k,l} \exp(2\pi i\cdot 
0.h(l)\cdot t)\exp (-2\pi i(L-1)\d^*_{k,l})|\Phi_k ,l\rangle $.

6. Apply to $\chi_6$ the transformation $\WH\ \U^q_{sec}\ \QFT_L$, then
 observe and discard ancilla.

The time of this algorithm is $2L+2t_{\QFT_L }+2t_{h}+w$ where $t_{\QFT_L}$ 
is the time of Fourier transform, $t_h$ is the time of enhancing accuracy and 
$w$ is the time of rotations.

\begin{Lemma}
Denote the result of algorithm 1-6 by $\U_{q,t}$. For any $\d >0$ for any 
$p, q, n, t$ altering so that $q=p+c, L=2^q$, $t: \ 0<t\leq CN$, where 
$c=\log_2 (\frac{4}{\d} )$, $p\geq c$, $C=\frac{\d}{14}$ we shall have
$$
\| \U_{q,t} -\U^t \| <\d
$$
\end{Lemma}

{\it Proof}

Introduce the simplifying notations: \newline
$\chi = \sum\limits_{k}\sum\limits_{l=0}^{L-1} y_{k,l} \exp (2\pi i \w_k t )
|\Phi_k ,l,h(l)\rangle$, 
$\tilde\chi = \QFT^{-1}_L \ (\U_{seq}^q)^{-1}\ \WH \exp(2\pi 
i\w_k t)\sum\limits_{k}
x_k |\Phi_k ,0^q \rangle$, \newline
$\chi' =\sum\limits_k \sum\limits_{l=0}^{L-1} y_{k,l} \exp(2\pi i \w_k t)\exp(
-2\pi i(L-1)\d^*_{k,l} )|\Phi_k ,l\rangle$.

By Lemma 1 $\|\chi_4 -\chi\| <\frac{\d}{2}$. The passages $\chi_4 \ar\chi_6$
and $\chi\ar\chi' $ are fulfilled by the same unitary transformation, which 
preserves lengths. Consequently, $\|\chi_6 -\chi' \| <\frac{\d}{2}$.
By Lemma 2 $\|\tilde\chi -\chi' \| <\frac{\d}{2}$. Then 
triangle inequality gives $\|\chi_6 -\tilde\chi\| <\d$. The passages $\chi_6
\ar U_{q,t}$ and $\tilde\chi \ar U^t$ are fulfilled by the same unitary
 transformation $\WH\ \U^q_{sec}\ \QFT_L$. 
Hence $\| U^t - U_{q,t} \| <\d$. Lemma 3
 is proved. 

By Lemma 3 this algorithm gives the prediction of state in time instant 
$O(N)$ in time $O(M)$ if classical algorithm enhancing accuracy obtain 
eigenvalues in negligible time. Thus if it is possible to enhance accuracy 
of eigenvalues from $\e_1$ to $\e$ the speedup will be
 $\frac{\e_1 (1-\rho )}{56\e}$, where $\rho$ is a fidelity, $0<\rho <1$.
The result can be formulated as 

\begin{Theorem}
Given a Hamiltonian of system with sparse spectrum and algorithm enhancing the 
accuracy of eigenvalues from $\varepsilon_1$ to $\varepsilon$ and a 
fidelity $\rho$, a state of the system in the time instant $\frac{1-
\rho}{14\ep}$
can be obtained in time $\frac{4}{(1-\rho )\ep_1}$ with this fidelity.
\end{Theorem}

\subsection{Generalizations}

\subsubsection{Sparse areas of spectrum}

A spectrum of real system like molecule typically contains strips where 
spectrum is continuos separated by gaps where energy levels are absent at
 all. Let $w$ be the maximal width of strips and $g$ be the minimal width 
of gapes. 

Regard the times of evolution less than $1/w$ (here we assume the system of
 units where Plank constant is unit). For these times the width of strips is
negligible and we can assume that $\ep_1 =1/g$ in Theorem 2. So in this
 situation $\ep =w$ and we obtain the following generalization of Theorem 2.

{\it If $t=o(1/w)$ we can predict the state $U(t)$ of primary system in time
 $t_{pre} = O(t\frac{w}{g} )$.}

Yet more generalization can be obtained if we consider the spectrum which is 
not sparse at all but has sparse areas. Here if the initial state of primary 
system is concentrated in sparse area, the prediction is possible with such
 probability which is equal to the degree of concentration. 

\subsubsection{Travelling in a time: predictions and restoration of a history}

A wizard transformation may be used not only for predictions a future but for
 restoration of a history as well. If we replace the time $t$ by $-t$ in 
predicting procedure then we obtain the state of primary system in time 
instant $-t$ which means the restoration of a history. Perform procedure 
from the section 4.2 with $-t$ instead of $t$, where $0<t<CN$. This makes 
possible to obtain the state of primary system in time of order $-N$ in the
 past. Thus we conclude the following generalization of Theorem 2.

\begin{Theorem}
Given a Hamiltonian of some system, its states in time instant $T=
\frac{1-\rho}{14\ep}$ in future as well as in time instant $-T$ in the past 
can 
be obtained in time $\frac{4}{(1-\rho )\ep_1}$ with fidelity $\rho$ provided
each
 eigenvalue can be quickly calculated with precision up to $\ep$ given its
 approximation with accuracy $\ep_1$.
\end{Theorem}

Note that for restoration of a history we don't need a sparse spectrum, may be
$\ep_1 =\ep$. This is surprising because the algorithm uses only operator $\U$
when the natural way to obtain a state from the past is to use an inverse
 transformation $\U^{-1}$ and there is no evident way to simulate the action
of the inverse transformation by means of $\U$.

\subsection{Examples}

\subsubsection{Shor factoring algorithm}

Formula (2) for a wizard simulation can be considered as a natural 
generalization for the quantum part of Shor factoring algorithm (\cite{Sh}).
Let a unitary operator $\U$ in (2) has the form $\U |x\rangle\ar |a x\ 
(\mod\ q)\rangle$, where $a x$ is a multiplication of integers,
$(a,q)=1$. Then eigenvectors of $\U$ have the form
$\frac{1}{\sqrt{r}}\sum\limits_{j=0}^{r-1} \exp (-2\pi i k j/r) |a^j \
(\mod\ q)\rangle$ and its eigenvalues are $\exp (2\pi i j/r)$ where $r$ is a period of $a$ (minimal integer such that $a^r \equiv 1\ (\mod\ q)$). If we apply an operator (2) with this $\U$ to zero initial state and observe the second register then by Lemma 1 we obtain an approximation of a number $j/r$ up to $O(1/N)$ with high probability. Using this procedure sequentially we can restore the value $r$ of a period. This procedure was used for factoring by Shor. 

Here we use a wizard simulation (2) as a technical element of the algorithm
4.2. Alternatively, one can try to use Kitaev's method instead of QFT in (2).
Namely, consider one controlling qubit and use Hadamard transform after
conditional application of $\U$ and repeat this with numerous controlling
qubits. Then by some auxiliary transformation it is possible to extract
eigenvalues in ancillary register (look at \cite{Ki} for details). But
then we must get rid of revealed eigenvalues like in the point 6 of the
 algorithm and it faces a little difficulty when using Hadamard transform
 instead of QFT in (2), because we have not $\U_{cond}^{-1}$ and must always 
manage with only $\U_{cond}$. 

\subsubsection{Grover search algorithm}

Consider Grover algorithm for the fast quantum search: 
$U(t)|\tilde 0\rangle=(I_{\tilde 0} I_a )^t |\tilde 0\rangle$, 
where $I_b$ denotes inversion 
of the state $b$: $I_b |b\rangle = -|b\rangle ,\ I_b |s\rangle = 
|s\rangle$ for $\langle b|s\rangle =0$. It was shown in \cite{Gr2}
 that if $|\langle\tilde 0 |a\rangle | =O(1/\sqrt{N} )$ 
then $U(t_1 )\approx |a\rangle$ for $t_1 =O(\sqrt{N} )$
independently of $\tilde 0$, where $N$ is the dimension 
of main space. What will happen if we apply Theorem 2 for 
$U$ as a primary evolution? The problem in fact will be two
 dimensional and the minimal gap between eigenvalues will be 
of order $1/\sqrt{N}$. Hence, by Theorem 2 we can predict the 
states $U(t)$ for $t=O(N)$ in time of order $\sqrt{N}$. As for 
$t=O(\sqrt{N})$ 
we obtain no additional speedup.

\subsection{About practical implementations}

The two main components of the algorithm are conditional iterations of a 
primary transformation $\U$ and QFT.

The main difficulty for the practical implementation of this method is in 
conditional iterations of $\U$. Given only a primary device realizing $\U$ 
one cannot immediately fulfil this iterations because it requires quantum
 control on the number of iterations. The solution may be following. 
Decomposition of a primary device into elementary parts that can be 
included to the quantumly controlled circuit and realize the conditional 
iterations of $\U$ for all parts simultaneously. All controlling qubits should 
be used in entangled state as a Shr\"oedinger cat. Given a quantum gate array 
computing $\U$ we can easily construct a new gate array computing $\U_{cond}$. 
To put it otherwise a conditional application of $\U$ is possible through the 
control on microscopic level (this bears a resemblance with the control in 
living cells). 

This scheme can be reformulated by means of analogous quantum computing if we 
consider QFT on an ancillary register as a passage to the canonically conjugate 
magnitude. Say, if we use a value of coordinates in operations with a register 
then canonically conjugate will be the corresponding impulse. The passage from the coordinate representation of a wave function to the impulse representation 
in one-dimensional case can be defined as 
$$
\phi (p)=\int\limits_{-\infty}^{+\infty} \exp (-ipx/h )\psi (x) dx
$$
where a probability to obtain impulse in a segment $(p,p+dp)$ is $|\phi (p)|^2 
dp/2\pi h$. Assume that we have one particle which can be located in $M$ points 
of the form $\ x=0,\frac{1}{M} ,\frac{2}{M},\ldots ,\frac{M-1}{M}$. Then the 
coordinate quantum space for one particle will be $M$ dimensional. Consider the 
corresponding integral sum for $\phi (p)$ in the system of units where Plank 
constant $h$ is one. This integral sum will be just the sum in the definition of 
QFT where $x$ plays a role of $s/M$ and $p/2\pi$ plays a role of $l$.

Then the main algorithm acquires the following general form. 

a). Primary evolution quantumly controlled by a magnitude containing in the 
properly prepared ancillary register. 

b). Simple actions depending on the canonically conjugate magnitude.

c). Repetition of a). 

Procedure of such a kind can predict a behavior of a primary system with 
sparse spectrum and restore the state of arbitrary system in the past. 
This scheme seems to be very simple and would be interesting to find its 
natural physical analog. The good starting point here may be the comparison 
between spectral features of the known systems and their functions and 
complexity. 

\section{Conclusion}

Formulate again the main result: the method is presented which makes 
possible to obtain states of a primary quantum system earlier than they 
appear naturally in its evolution provided the spectrum of system is sparse.
 This speedup will be the more if the gaps between continuous strips of 
spectrum increase comparatively to the width of strips. This method can be
 applied also for restoration of states of the arbitrary primary system in the 
past.
 This method yields a speedup and it can make possible to fit into the time 
when coherent states exist and thus fight decoherence in quantum computations.

\section{Acknowledgements}

I am grateful to Lov Grover for his kind invitation to Bell Labs, discussions
 and attention to my work. I also thank Peter Hoyer and David DiVincenzo for 
fast replies and useful comments.


\begin{thebibliography}{99}
\bibitem[AL]{AL} D.S.Abrams, S.Lloyd, A quantum algorithm providing 
exponential speed increase for finding eigenvalues 
and eigenvectors {\it lanl e-print quant-ph/9807070}
\bibitem[BBCMW]{BBCMW} R. Beals, H. Buhrman, R.Cleve, M. Mosca, R. De Wolf,
 Tight Quantum Bounds by Polynomials, {\it lanl e-print quant-ph/9802049}
\bibitem[FGGS]{FGGS} E. Farhi, J. Goldstone, S. Gutmann, M. Sipser, A 
limit on the Speed
of Quantum Computation in Determining Parity, {\it Phys. Rev. Lett., 81,
1998, 5442-5444, lanl e-print quant-ph/9802045}
\bibitem[Gr]{Gr} L. K. Grover,  A fast quantum mechanical algorithm for
database search. {\it Proceedings, STOC 1996, 212-219. Philadelphia PA USA, 
lanl e-print quant-ph/9605043}
\bibitem[Gr2]{Gr2} L.K.Grover, Rapid sampling through quantum computing. 
{\it lanl e-print 
quant-ph/9912001}
\bibitem[Ki]{Ki} A. Yu. Kitaev, Quantum measurements and the Abelian 
Stabilizer Problem {\it lanl e-print quant-ph/9511026}
\bibitem[Oz]{Oz} Y.I.Ozhigov, Quantum computers speed up classical
with probability zero {\it Chaos, Solitons and Fractals, 10(10), 1999, 
1707-1714, lanl e-print quant-ph/9803064}
\bibitem[Sh]{Sh} P. W. Shor, Polynomial-time algorithms for prime 
factorization and discrete logarithms on a quantum computer, 
{\it SIAM J. Comp. 1997,
26, No. 5, 1484-1509, lanl e-print quant-ph/9508027}
\end{thebibliography}
\end{document}